\documentclass[twocolumn]{aa}
\usepackage{graphicx}
\usepackage{natbib}
\usepackage{txfonts}
\def\cm-2{cm$^{-2}$}

\def\ros{{\it ROSAT}}
\def\chandra{{\it Chandra}}

\def\xmm{{XMM-Newton}}
\def\n253{\object{NGC~253}}
\def\ulx{\object{NGC~253~ULX1}}
\def\M101{\object{M101~ULX-1}}

\def\spose#1{\hbox to 0pt{#1\hss}}
\def\gaeq{\mathrel{\spose{\lower 3pt\hbox{$\mathchar"218$}}
     \raise 2.0pt\hbox{$\mathchar"13E$}}}

\newcommand{\ergs}[1]{$\times10^{#1}$ \hbox{erg s$^{-1}$}}
\newcommand{\oergs}[1]{$10^{#1}$ erg s$^{-1}$}
\newcommand{\hcm}[1]{$\times10^{#1}$ cm$^{-2}$}
\newcommand{\ohcm}[1]{$10^{#1}$ cm$^{-2}$}

\begin{document}
   \title{The recurrent ultra-luminous X-ray transient \ulx\thanks{Based on observations obtained with XMM-Newton, an ESA science mission with instruments and contributions directly funded by ESA Member States and NASA}}


   \author{M. Bauer
          \and
          W. Pietsch
          }


   \institute{Max-Planck-Institut f\"ur extraterrestrische Physik, Giessenbachstra\ss e, 85741 Garching, Germany\\
              \email{mbauer@mpe.mpg.de}
             }

   \date{Received 22 June 2005; Accepted 28 June 2005}

   \abstract{
   We present the results of \ros\ and \xmm\ observations of the recurrent ultraluminous X-ray source (ULX) \ulx.
   This transient is one of the few ULXs that was detected during several outbursts. 
   The luminosity reached 1.4\ergs{39} and 0.5\ergs{39} in the detections by \ros\ and \xmm, respectively, indicating a black hole X-ray binary (BHXRB) with a mass of the compact object of $\ge$11 M$_{\odot}$.
   In the \ros\ detection \ulx\ showed significant variability, whereas the luminosity was constant in the detection from \xmm.
   The \xmm\ EPIC spectra are well-fit by a bremsstrahlung model (kT$=2.24\,$keV, $N_H=1.74$\hcm{20}), which can be used to describe a comptonized plasma.
   No counterpart was detected in the optical I, R, B, NUV and FUV bands to limits of 22.9, 24.2, 24.3, 22 and 23 mag, respectively, pointing at a XRB with a low mass companion.
      
   }

   \maketitle
%

\section{Introduction}

Ultra-luminous X-ray sources (ULXs) are extra-nuclear compact X-ray sources with luminosities considerably exceeding the Eddington luminosity for stellar mass X-ray binaries of $\sim 2$\ergs{38} \citep{2000ApJ...535..632M}.
However their luminosities are still lower than that of active galactic nuclei (AGN).

There are currently four preferred models to explain the luminosities of these objects. 
The first is that ULXs are intermediate mass black holes (IMBHs: M$_\mathrm{BH}\sim 10^2-10^5 \mathrm{M}_\odot$).
However IMBHs are at present not explainable with stellar evolutionary models.
The alternatives are stellar-mass black hole X-ray binaries where either photon bubble instabilities allow super-Eddington luminosities \citep{2002ApJ...568L..97B}, anisotropically emitting X-ray binaries \citep{2001ApJ...552L.109K}, or that ULXs are micro-quasars that are observed down the beam of their relativistic jet \citep[e.g.][]{1997MNRAS.286..349R}.
It is therefore important to increase the sample of ULXs to find arguments that favour or exclude the above models.
One attempt was the search for ULXs in 313 nearby galaxies from \ros\ HRI observations by \citet[ hereafter LB2005]{2005ApJS..157...59L}.
A target of this search was the starburst galaxy \n253\ where they found 21 X-ray sources but only one of them matched their criteria for an ULX (\ulx).
This source is located within, but close to the north-east boundary of the D25 ellipse of \n253.

We here report on a more detailed analysis of \ulx\ including \ros, \xmm\ and \chandra\ data, and specifically on the detection of a second outburst in one of the \xmm\ observations.

\section{Search for the source in XMM-Newton, Chandra and ROSAT archives}

We searched the \ros, \chandra\ and \xmm\ archive for observations of \n253.
The results are listed in Table \ref{ObsTab}.
Except for two \chandra\ observations the position of \ulx\ was always in the field of view (FOV). 
Besides the first detection in \ros\ observation 601111h (LB2005), \ulx\ was only visible in \xmm\ observation 0110900101.
These \xmm\ and \ros\ HRI data are further discussed in Sect. 3 and 4, respectively.

For the remaining observations we determined $3\sigma$ upper limits for the count rate.
From that we obtained upper limit for fluxes and luminosities (cf Table \ref{ObsTab}).
We used WebPIMMS (v3.6c) with the spectral model we got from the analysis of \xmm\ observation 0110900101 to determine energy conversion factors.
The long term light curve of \ulx\ is shown in Fig. \ref{LcFig}. 

\begin{table*}
\begin{minipage}[t]{\textwidth}
\caption{Individual observations of \ulx.}
\label{ObsTab}
\centering
\renewcommand{\footnoterule}{}  
\begin{tabular}{lccrrr}
\hline \hline
Date& Instrument & Observation ID & Duration & \multicolumn{1}{c}{Flux$^a$} & \multicolumn{1}{c}{$L_\mathrm{X}$\footnote{$0.3-10\,$keV luminosity assuming a distance of 2.58 Mpc and a bremsstrahlung model (kT$=2.24\,$keV, $N_H=1.74$\hcm{20})}} \\
    &            &                & \multicolumn{1}{c}{(ks)}     & \multicolumn{1}{c}{(erg cm$^{-2}$ s$^{-1}$)}   & \multicolumn{1}{c}{(\oergs{39})}\\
\hline
 1991-12-08 & \ros  	& 600088h-0	& 3.1	& $<1.1\times 10^{-13}$ & $<0.08$\\
 1991-12-25 & \ros  	& 600087p-0	& 11.6	& $<2.2\times 10^{-14}$ & $<0.02$\\
 1992-06-03 & \ros  	& 600087p-1	& 11.2	& $<4.4\times 10^{-14}$ & $<0.03$\\
 1992-06-05 & \ros  	& 600088h-1	& 25.7	& $<4.4\times 10^{-14}$ & $<0.03$\\
 1995-01-03 & \ros 	& 600714h	& 11.0	& $<5.2\times 10^{-14}$ & $<0.04$\\
 1995-06-13 & \ros 	& 600714h-1	& 19.8	& $<2.0\times 10^{-14}$ & $<0.02$\\
 1997-12-20 & \ros 	& 601111h	& 17.5	& $1.8\times 10^{-12}$ &  1.4\\
 1998-07-01 & \ros 	& 601113h	& 2.0	& $<2.8\times 10^{-13}$ & $<0.2$\\
 1999-12-16 & \chandra	& 969		&	\multicolumn{3}{c}{not in FOV}\\
 1999-12-27 & \chandra	& 790		&	\multicolumn{3}{c}{not in FOV}\\
 2000-06-03 & \xmm 	& 0125960101	& 39.2	& $<4.3\times 10^{-14}$ & $<0.03$\\
 2000-06-04 & \xmm 	& 0125960201	& 14.2	& $<1.8\times 10^{-14}$ & $<0.01$\\
 2000-08-16 & \chandra 	& 383		& 2.16	& $<8.8\times 10^{-15}$ & $<0.007$\\
 2000-12-14 & \xmm 	& 0110900101	& 29.6	& $6.3\times 10^{-13}$  &  0.5\\
 2003-06-19 & \xmm 	& 0152020101	& 110.5	& $<3.6\times 10^{-15}$ & $<0.003$\\
 2003-09-19 & \chandra 	& 3931		& 83.6	& $<4.4\times 10^{-15}$ & $<0.003$\\
\hline
\end{tabular}
\end{minipage}
\end{table*}

\section{Detailed analysis of \xmm\ observation 0110900101}

\ulx\ was detected for the second time on 2000 December 14 with \xmm.
The position of the source was within the FOV of both of the MOS and the PN cameras.
We used the latest version of the Science Analysis System (SAS v6.1) to process the obtained data from the PN (thin filter) and the two MOS (medium filter) detectors. 
Periods of high background were determined and excluded from further analysis. 
The low background exposure times for PN and for the MOS instruments were 23.0 ks and 24.4 ks, respectively.

We applied the source detection tasks {\tt eboxdetect} and {\tt emldetect} only on the
data from the PN detector, as the source was positioned far from the optical axis and close to the edge of the FOV on the MOS detectors.
The obtained position was then corrected using optical reference coordinates from the USNO B1 catalogue \citep{2003AJ....125..984M} of three AGN, identified by \citet[ sources X4, X22, X58]{1999A&A...342..101V}. 
The corrected position in J2000 coordinates is $\alpha =$ 00h48m20.11s, $\delta = -25^\circ 10\arcmin 10\farcs4$ with an error in position of 0\farcs3.
The derived position is well within the positional errors given by LB2005 for \ulx.

A foreground star with a B magnitude of $\sim 13$ \citep{2003AJ....125..984M} is located close (15.5\arcsec) to the obtained position (Fig. \ref{OptFig}). 
We can rule out that the actual detection of \ulx\ in observation 0110900101 was caused by this star, as its proper motion of -9.2$\,$mas/yr in RA$\cdot \cos($DEC$)$ and -3.6$\,$mas/yr in DEC \citep{2004AJ....127.3043Z} is too small to match the detected position of \ulx\ with that of the star within the period of observations. 
Additionally there was no detection of the source in other \xmm\ observations using the same filter.

We extracted energy spectra for \ulx\ for all EPIC detectors.
For the PN chip we included source counts from an elliptical region with major and minor axes of 27.6\arcsec\ and 12.3\arcsec\ respectively. 
The background region was a circular source-free region with a radius of 48\arcsec\ on the same CCD close to the source.
For MOS the source extraction region was an ellipse with major (minor) axes of 28.95\arcsec\ (11.3\arcsec) for MOS1 and 31.35\arcsec\ (15.15\arcsec) for MOS2, respectively.
The background regions were circles with radii of 68\arcsec\ and 80\arcsec\ for MOS1 and MOS2, respectively.
After subtracting the background the spectra for each instrument were rebinned to a significance level of $3\sigma$.

\begin{figure}
  \centering
  \resizebox{8.8cm}{!}{\includegraphics[angle=-90,clip]{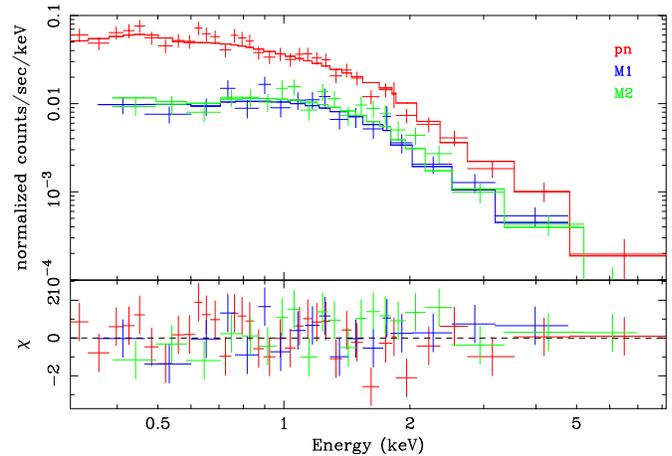}}
  \caption{Comparison of the PN and MOS spectra of \ulx\ with the best-fit bremsstrahlung model. In the lower panel the residuals (in units of $\sigma$) between data and model are shown.}
  \label{SpecFig}
\end{figure}

\begin{table}
\begin{minipage}[t]{8.8cm}
\caption{Models for the source spectrum of \ulx}
\label{FitTab}
\centering
\renewcommand{\footnoterule}{}  
\begin{tabular}{lcr@{=}lc}
\hline \hline
Model\footnote{po: power law, bremss: thermal bremsstrahlung, mekal: Mekal thermal plasma, diskbb: multiple blackbody disk, diskpn: accretion disk around a black hole} 	& $N_H$\footnote{all models are modified by foreground absorption (XSPEC model {\tt wabs}).} 		&\multicolumn{2}{c}{Model parameter} 		& $\chi^2_{red}$\\
	& (\ohcm{20})		&\multicolumn{2}{c}{}				&		\\
\hline
\smallskip
po 	&1.30			&$\Gamma$ & $\,1.94\pm0.05$		&1.663	\\
\smallskip
bremss 	&$1.74^{+0.02}_{-0.01}$	&kT & $\,2.24^{+0.38}_{-0.31}\,$keV	&0.961	\\
\smallskip
mekal	&1.30			&kT & $\,3.17\pm 0.19\,$keV		&2.042	\\
\smallskip
diskbb 	&1.30			&kT & $\,0.62\pm0.04\,$keV		&1.671	\\
\smallskip
diskpn	&1.30			&kT & $\,0.69^{+0.06}_{-0.07}\,$keV	&1.366	\\
	&			&$R$ & $\,5.34_{-2.34}^{+9.79}\,R_S$&	\\
\hline
\end{tabular}
\end{minipage}
\end{table}

For the spectral analysis XSPEC 11.3.1 was used. 
The best-fit parameters from different models provided within XSPEC are listed in Table \ref{FitTab}.
Using the PN and MOS spectra simultaneously the source spectrum was best fitted with a bremsstrahlung model (Fig. \ref{SpecFig}). 
The fit of the multicolour disk blackbody model (diskpn) would also be acceptable.
However, we favour the bremsstrahlung model since it is less complex and gives a better $\chi^2_{red}$.
Except for the bremsstrahlung model the foreground absorption ($N_H$) had to be fixed to the Galactic foreground absorption as a lower limit  \citep[1.30\hcm{20}, ][]{1990ARA&A..28..215D}. 
If the parameter was free to adjust it converged to unreasonably low values.

From the best fitting spectral model we calculated the source flux and, assuming a distance of 2.58 Mpc \citep{1991AJ....101..456P} we derived an unabsorbed luminosity of 5.0\ergs{38} in the 0.3-10.0 keV band.

In order to study the temporal behaviour of the source a background corrected light curve was created using the tasks {\tt evselect} and {\tt lccorr}.
The source count rate was constant at about 0.8 ct s$^{-1}$ within the errors of about 15\% during observation 0110900101.

\section{Analysis of ROSAT observation 601111h}

\begin{figure}
  \centering
  \resizebox{8.8cm}{!}{\includegraphics[angle=-90,clip]{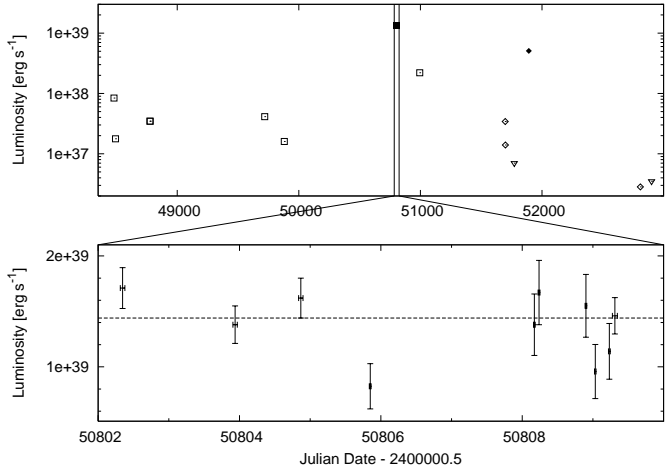}}
  \caption{Light curve of \ulx. Upper panel: solid symbols represent detections, open symbols 3$\sigma$ upper limits of \ulx. Different instruments are represented by different symbols: \ros\ (squares), \xmm\ (diamonds) and \chandra\ (triangles). Lower panel: Single \ros\ HRI exposures with errorbars from observation 601111h where the source was detected. The length of each observation is indicated by the x-errorbar. In contrast to the upper panel the lower panel plot is linear in luminosity.}
  \label{LcFig}
\end{figure}

The first detection of \ulx\ was in \ros\ observation 601111h (LB2005).
The observation (total exposure time 17.5 ks) is spread over ten observing intervals, with different exposure and waiting time for the individual observations. 
The source was bright enough to determine luminosities for each of these observation intervals.

We calculated count rates using the EXSAS source detection task {\tt detect/sources}. 
To reduce noise we only analysed HRI channel 2-15.
We used WebPIMMS (v3.6c) and the spectral model retrieved from the \xmm\ observation (bremsstrahlung, kT$=2.24\,$keV, $N_H=1.74$\hcm{20}) to determine energy conversion factors to obtain the corresponding fluxes and luminosities (see lower panel of Fig. \ref{LcFig}).
The luminosity averaged over the whole observation (1.43\ergs{39}) is indicated by the dashed line.

During the observation the source showed significant variability by at least a factor of 2.

\begin{figure}
   \centering
   \resizebox{8.8cm}{!}{\includegraphics[clip]{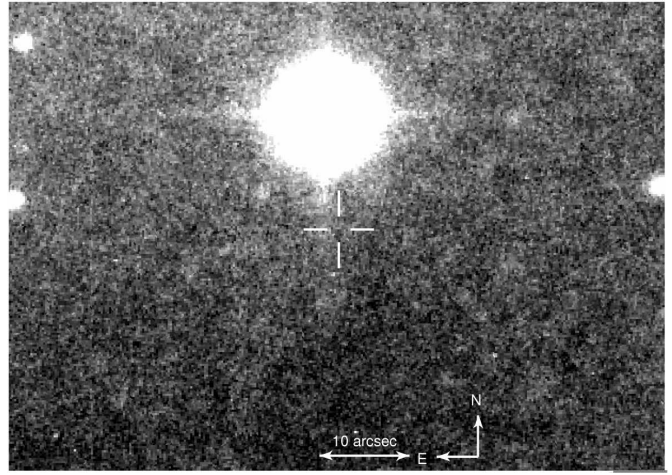}}
   \caption{R-Band optical image taken with the Wide Field Imager on the MPG-ESO 2.2m Telescope. The source is located close to a $\sim 13\,$mag star. The R magnitude is 24.2.}
   \label{OptFig}
\end{figure}

\begin{figure}
   \centering
   \resizebox{8.8cm}{!}{\includegraphics[clip]{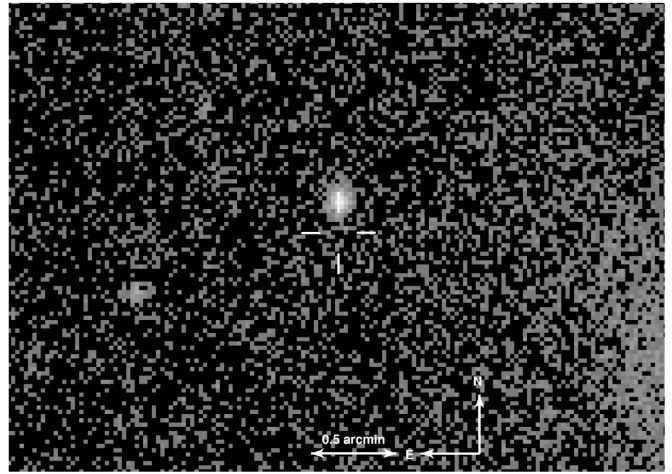}}
   \caption{Near UV image taken with GALEX. The NUV magnitude is 22.}
   \label{GalexFig}
\end{figure}

\section{Discussion}

We detected the recurrence of \ulx\ in the \xmm\ observation from 2000 December 14. 
This was the first detection after the outburst in 1997, reported from \ros\ HRI observations by LB2005. 
In all other observations of \n253\ the luminosity of the source was below the detection limit. 
This implies brightness variability by at least a factor of 500.
Its fastest change in luminosity $(\mbox{L}_{\mbox{max}}/\mbox{L}_{\mbox{min}})$ exceeds a factor of 71 in 120 days.

The improved position (errors of 0\farcs 24 compared to 4\arcsec - 10\arcsec) of the source determined in Sect. 3 allowed us to search for optical counterparts.
We checked images taken with the Wide Field Imager (WFI) on the MPG-ESO 2.2m Telescope at La Silla in the R- (Fig. \ref{OptFig}), I- and B-band (limiting magnitudes 24.2, 22.9 and 24.3, respectively) and images taken with the Galaxy Evolution Explorer (GALEX, a space telescope from NASA observing in the ultraviolet) in the NUV (Fig. \ref{GalexFig}) and FUV (limiting magnitudes 22 and 23, respectively), but no counterpart could be detected.

With the data discussed in Sect. 3 and 4 we can exclude that \ulx\ is either a foreground object or a background AGN based on three arguments: 
(i) We estimated the $\log(f_{\mbox{x}}/f_{\mbox{opt}})=\log f_{\mbox{x}}+(m_{\mbox{v}}/2.5)+5.37>3.2$ using the flux of the \ros\ detection and a lower limit for $m_{\mbox{v}}$ of 24.2 (averaging the limiting magnitudes of the R- and the B-WFI images, see above). 
Following \cite{1988ApJ...326..680M} this value exceeds that expected for galactic sources ($-4.6$ to $-0.6$) as well as AGNs ($-1.2$ to $+1.2$).
(ii) The variability of \ulx\ is by a large factor higher than the typical value observed for AGNs $(\sim10-60)$.
(iii) \ulx\ shows a bremsstrahlung spectrum, whereas spectra of AGNs above $2\,$keV are typically fitted by a power law.
The recurrent outbursts also exclude that the source is the luminous remnant of a recent supernova, like e.g. SN1993J in M81 \citep{2003A&A...406..969Z}.

The X-ray spectrum may indicate that \ulx\ is a low mass X-ray binary (LMXB). 
The X-ray emission in these objects is created in the optically thin boundary layer between the disk and the neutron star and comptonization may dominate the spectral emission \citep{1988ApJ...324..363W}, leading to a spectrum that can be fitted with a bremsstrahlung model.
However the \ros\ HRI peak luminosity of 1.43 \ergs{39} is very high for typical LMXBs.
Other systems that show bremsstrahlung spectra are black hole XRBs, e.g. \object{Cyg X-1} \citep{1979Natur.279..506S}, \object{LMC X-3} and \object{X1755-33} \citep{1988ApJ...324..363W}.
These systems may contain a high or low mass companion.

An additional argument for a low mass companion comes from the lack of an optical counterpart (see above).
High mass X-ray binaries (HMXBs) should be detectable at about 22 to 24 mag, extrapolating V magnitudes from HMXBs in the Magellanic Clouds \citep{2000A&AS..147...25L}.
We would have detected an object of this brightness in the WFI data.

The luminosity of a compact object radiating at the Eddington limit is given as $L_{Edd}=1.5\times10^{38}(M/M_\odot)$ \hbox{erg s$^{-1}$}, when electron scattering dominates the opacity. 
Luminosities higher than 2\ergs{38} (corresponding to a $1.4M_\odot$ object, commonly assumed as the maximum mass of a neutron star) suggests that the compact object is a black hole.
According to \ulx's maximum luminosity of 1.43\ergs{39} the lower limit for the mass of the black hole is 11 M$_\odot$.
Therefore \ulx\ is not required to be an IMBH.

Another argument against an IMBH is the temperature of \ulx.
\cite{2004ApJ...614L.117M} compared intermediate mass black hole candidate ULXs and stellar mass black holes with respect to luminosity and temperature.
If we assume the multicolour disk blackbody model then \ulx's position in the luminosity-disk temperature diagram \citep[Fig. 2 in][]{2004ApJ...614L.117M} indicates that \ulx\ is not an IMBH, but a stellar mass black hole.

Recently another object was found that, like \ulx, showed also a bremsstrahlung spectrum: X-44 in the Antennae Galaxies (NGC 4038/4039) \citep{2004ApJ...609..728M}.
The temperature of X-44 is $3.7\pm0.5\,$keV and its luminosity is $1.0^{+1.3}_{-0.2}$ \ergs{40}.
This temperature is about a factor of 1.5 higher than in \ulx, and the luminosity exceeds the luminosity of \ulx\ by a factor of 15 compared to the outburst in 1997.

Another interesting ULX to compare \ulx\ with is \M101\ \citep{0503465}.
It was the first ULX that like \ulx\ has been observed during more than one ultra-luminous outburst.
Like many other ULXs the spectrum of \M101\ is best described with an absorbed blackbody model, but the temperature of $\sim 50-160\,$eV is rather low.
\M101\ has a peak luminosity of about \oergs{41} ($0.3-7\,$keV), and the hardness of its spectrum changed between different observations.
We do not know whether the spectrum of \ulx\ changed in the two observations, as the \xmm\ data provided the very first spectrum of the source.
During 12 years of observations \ulx\ showed two outbursts with an interval of three years.
In \M101\ the two outbursts are only seperated by half a year.
On shorter time scales \ulx\ showed only one drop in luminosity by a factor of $\sim 2$ during the \ros\ observation, and in the \xmm\ observation (exposure time $8.2\,$h) no variability could be detected.
\M101\ on the other hand does show short-time-scale variability. 
Its luminosity changed by a factor of $\gaeq 10$ on a time scale of hours.
The lack of short time variability of \ulx\ argues against the relativistic beaming model, since this would require a very stable jet \citep{1997MNRAS.286..349R}.

\begin{acknowledgements}
We thank G. Szokoly for providing us with the images of \n253 from the Wide Field Imager, which are based on observations made with ESO Telescopes at the La Silla and Paranal Observatory.
Some of the data presented in this paper were obtained from the Multimission Archive at the Space Telescope Science Institute (MAST). STScI is operated by the Association of Universities for Research in Astronomy, Inc., under NASA contract NAS5-26555. Support for MAST for non-HST data is provided by the NASA Office of Space Science via grant NAG5-7584 and by other grants and contracts. 
The \xmm\ and the \ros\ project is supported by the Bundesministerium f\"ur Bildung und Forschung/Deutsches Zentrum f\"ur Luft- und Raumfahrt (BMBF/DLR), the Max-Planck Society and the Heidenhain-Stiftung.
M.B. acknowledges support from the International Max Planck Research School on Astrophysics (IMPRS).
\end{acknowledgements}

\bibliographystyle{aa}
\bibliography{/home/mbauer/work/Publications/papers}

\end{document}